\documentclass{aa}
\usepackage{mathabx}
\usepackage{tabularx}
\usepackage{natbib}
\bibpunct{(}{)}{;}{a}{}{,}

\begin{document}

%PREAMBLE

\title{The curious case of Mars' formation}
\author{Jason Man Yin Woo\inst{1,2}, Ramon Brasser\inst{1,6}, Soko Matsumura\inst{3}, Stephen J. Mojzsis\inst{4,5,6}, Shigeru Ida\inst{1,2}}

\offprints{J. M. Y. Woo, \email{jasonwoohkhk@gmail.com}}

\institute{Earth Life Science Institute, Tokyo Institute of Technology, Meguro-ku, Tokyo, 152-8550, Japan
\and Department of Earth and Planetary Sciences, Tokyo Institute of Technology, Meguro-ku, Tokyo, 152-8550, Japan
\and School of Science and Enginerring, Division of Physics, Fulton Building, University of Dundee, Dundee DD1 4HN, UK
\and Department of Geological Sciences, University of Colorado, UCB 399, 2200 Colorado Avenue, Boulder, Colorado 80309-0399, USA
\and Institute for Geological and Geochemical Research, Research Center for Astronomy and Earth Sciences, Hungarian Academy of Sciences, 45 Buda$\ddot{\rm o}$rsi Street, H-1112 Budapest, Hungary
\and Collaborative for Research in Origins (CRiO), The John Templeton Foundation - FfAME Origins Program}

\abstract{Dynamical models of planet formation coupled with cosmochemical data from martian meteorites show that Mars’ isotopic composition is distinct from that of Earth. Reconciliation of formation models with meteorite data require that Mars grew further from the Sun than its present position. Here, we evaluate this compositional difference in more detail by comparing output from two $N$-body planet formation models. The first of these planet formation models simulates what is termed the `Classical' case wherein Jupiter and Saturn are kept in their current orbits. We compare these results with another model based on the ‘Grand Tack’, in which Jupiter and Saturn migrate through the primordial asteroid belt. Our estimate of the average fraction of chondrite assembled into Earth and Mars assumes that the initial solid disk consists of only sources of enstatite chondrite composition in the inner region, and ordinary chondrite in the outer region. Results of these analyses show that both models tend to yield Earth and Mars analogues whose accretion zones overlap. The Classical case fares better in forming Mars with its documented composition (29\% to 68\% enstatite chondrite plus 32\% to 67\% ordinary chondrite) though the Mars analogues are generally too massive. However, if we include the restriction of mass on the Mars analogues, the Classical model does not work better. We also further calculate the isotopic composition of $^{17} \rm O$, $^{50} \rm Ti$, $^{54} \rm Cr$, $^{142} \rm Nd$, $^{64} \rm Ni$, and $^{92} \rm Mo$ in the martian mantle from the Grand Tack simulations. We find that it is possible to match the calculated isotopic composition of all the above elements in Mars' mantle with their measured values, but the resulting uncertainties are too large to place good restriction on the early dynamical evolution and birth place of Mars.}


\titlerunning{The curious case of Mars' formation}
\authorrunning{Woo et al.}
\maketitle

%Introduction
\section{Introduction}
\label{sec:intro}

Basic details of the physiochemical dynamics which led to the formation of the terrestrial planets have remained elusive until relatively recently. Moore’s Law improvements in computational performance have transformed the way dynamicists are able to study the physical and chemical processes of terrestrial planet accretion via increasingly sophisticated numerical simulations. Analyses of terrestrial rocks and meteorites are being made with ever-more powerful analytical tools that yield data at unprecedented precision and accuracy. We now, therefore, have a firmer understanding of the chemical and isotopic compositions of Earth, the Moon, Mars, Vesta, various asteroidal meteorites derived from some 100-150 parent bodies, and some comets. These data help to constrain both the absolute and relative chronology of planet formation that is subsequently fed into modern dynamical models of accretion  \citep[e.g.][]{morbidelli2012building}.

Apart from Earth, Mars is the only other planet for which we have direct samples. Geochemical and isotopic analysis of these samples allows us to narrow the uncertainties of the timing of Mars’ formation. For example, \cite{dauphas2011hf} analysed a suite of martian meteorites to argue that Mars reached about half of its current size in approximately 1.8 Myr. Mars likely finished its accretion within $\sim$10 Myr of the start of the solar system. It is straight forward to think that Earth requires more time to form because of its much larger mass. Indeed, derivation of timescales for terrestrial accretion from the Hf-W chronometer supports this idea by showing that Earth finished its core formation at least 30 Myr after the Calcium-aluminium-rich inclusion (CAI) formation (e.g. \cite{kleine2009hf} and references therein). These observations can be summarised as a general time line for the formation of the terrestrial planets, and combining such geochemical data with numerical simulations helps in efforts to sort out the mystery of the formation of the sampled terrestrial planets and by extension the whole of the inner solar system.

\section{Background}
\subsection{The final stage of planet formation}
\label{subsec:final_form}
According to traditional dynamical models of the late stage of planet formation, coagulation of planetesimals into planetary embryos and subsequent giant impacts between embryos gave rise to the terrestrial planets. Several different models have been proposed for this scenario. The foundational one is the Classical model, in which all the terrestrial planets and the two gas giants formed near their current locations. \cite{chambers2001making} performed simulations with different masses of planetesimals and embryos to study the growth of terrestrial planets with Jupiter and Saturn on their current orbits. Simulations showed that it is typical to produce three to four terrestrial planets with orbital characteristics similar to the current system. A recurring problem with the simulations, however, is that the masses of computed planets which form close to 1.5 AU are consistently greater than the current mass of Mars. This was recapitulated in simulations by \cite{raymond2009building}, especially when both Jupiter and Saturn were provided circular orbits. To avoid forming a massive Mars, an alternative solution was proposed by \cite{agnor1999character} and expanded upon by \cite{hansen2009formation}; this work initially confined all solid material to the area in between the current locations of Venus and Earth at 0.7 AU and 1 AU in an annulus model. With this setting, \cite{hansen2009formation} successfully reproduced the mass-distance relation of the terrestrial planets. If all the material making up the terrestrial planets was confined within a small annulus, what is the underlying mechanism that leads to this configuration? The Grand Tack model proposed by \cite{walsh2011low} provides a reasonable explanation.

In the Grand Tack, Jupiter first formed in a position closer to the Sun than its current orbit and opened a gap in the gas disk. Owing to the unbalanced tidal torque acting on the planet, Jupiter migrated inwards through type-II migration \citep{lin1986tidal}. At the time Jupiter was migrating, Saturn formed further away and slowly accreted its gaseous envelope. Saturn then began migrating inwards once it reached about 50 Earth masses, caught up with Jupiter, and the two gas giants were trapped in the 2:3 mean motion resonance \citep{masset2001reversing}. With this specific orbital spacing and mass ratio between the two gas giants, the total torque on both planets reversed and therefore the migration direction also reversed outward from the Sun. The tack location (the location where Jupiter reversed its direction of migration) is set to be 1.5 AU in order to confine the growth of Mars \citep{walsh2011low}. This mechanism shows how the migration of Jupiter reshaped the structure of the inner solid disk. Two possible scenarios emerge from this model: either Jupiter scattered the planetesimals and embryos in its path away, or pushed them into the inner region by trapping them in mean motion resonance. In either case, when Jupiter reversed its migration, the region in between 1 and 1.5 AU was cleared. On the other hand, solid material piled up within 1 AU, thus duplicating the outer edge configuration suggested by the annulus model. Therefore, the mass-distance relation of the terrestrial planets and the low mass of Mars are explicable by the Grand Tack model. 

Another possible mechanism creating an annulus near the current orbits of the terrestrial planets is proposed by \cite{drkazkowska2016close}. In this model, pebbles pile up in radial distribution due to a combination of growth and radial drift of dust and therefore a narrow annulus of planetesimals close to 1 AU can be formed through the streaming instability \citep{johansen2007rapid,ida2016formation}. This model implies that all the terrestrial planets form from the same building blocks and therefore that (see below) their bulk and isotopic composition should be similar to each other. However, the isotopic compositions suggest that this may not be the case.   

\subsection{Bulk compositional differences between Earth and Mars}
\label{subsec:comp_diff}
The bulk isotopic compositions of Earth and Mars are different. The strongest evidence for this is the three oxygen isotope system (expressed as $\Delta^{17} \rm O_{\rm VSMOW}$ = $\delta^{17} \rm O_{\rm VSMOW}$ -- 0.52$\delta^{18} \rm O_{\rm VSMOW}$, where $\delta$-notation denotes deviations in parts-per-thousand where isotopic ratios are normalised to the standard mean ocean water and $\Delta$ expresses the deviation of these coupled ratios from the terrestrial mass-fractionation line of slope $m$ = 0.52). Bulk Mars is enriched in the minor oxygen isotope ($^{17} \rm O$) with respect to terrestrial and lunar values \citep[e.g.][]{franchi1999oxygen,rubin2000angeles,mittlefehldt2008oxygen,agee2013unique,wittmann2015petrography}. Other isotopic systems that trace nucleosynthetic anomalies, such as Titanium ($\varepsilon^{50} \rm Ti$), Chromium ($\varepsilon^{54} \rm Cr$) and Nickel ($\varepsilon^{64} \rm Ni$), likewise show clear differences between average terrestrial and martian values \citep{brasser2017cool}. No known mass-dependent process can account for these deviations \citep{qin2016nucleosynthetic}; they are instead attributable to implantation of neutron-rich isotopes by nearby supernovae at the time of solar system coalescence \citep{qin2010contributors}. This nucleosynthetic contamination is widely assumed to have been more influential in the composition of outer regions of the disk. Although under debate, a number of studies have proposed that the heliocentric composition gradient of the solid disk can be mapped by the isotopic differences of nucleosynthetic tracers that were implanted in different locations within the planet-forming disk. Taken together, it has been proposed that isotopic compositional differences between Earth and Mars hint that they were formed with different mixtures of source components \citep{wanke1988chemical,wanke1994chemistry,lodders2000oxygen,warren2011stable}, with the further implication that source locations for the two worlds were different \citep[cf.][]{fitoussi2016building}.

The difference between the source location(s) for components that gave rise to Earth and Mars is lent further credence by several recent isotope studies in terrestrial samples and martian meteorites. Assuming that the feed-stock for accretion of both Earth and Mars was dominated by chondritic material, analysis suggests that Earth should accrete  about 70$\%$ of enstatite chondrite, about 25$\%$ of ordinary chondrite and less than 5$\%$ of carbonaceous chondrite in order to reproduce the well-documented isotopic compositions of $^{17} \rm O$, $^{48} \rm C$, $^{50} \rm Ti$, $^{62} \rm Ni$ or $^{64} \rm Ni$ , $^{92} \rm Mo$ and $^{100} \rm Ru$ \citep{dauphas2014calcium,dauphas2017isotopic,fischer2017ruthenium}. Mars, on the other hand, has a larger uncertainty in its bulk composition. Its bulk composition is suggested to be 45\% enstatite chondrite and 55\% ordinary chondrite to account for the planet's documented $^{17} \rm O$, $^{50} \rm Ti$, $^{54} \rm Cr$, $^{62} \rm Ni$ and $^{92} \rm Mo$ isotopic values as determined from the martian meteorites \citep{sanloup1999simple,tang201460fe}. With a more rigorous analysis of uncertainties, \cite{brasser2018GRL} recorded 29\% to 68\% entatite chondrite and 32\% to 67\% ordinary chondrite in their analysis, which is in agreement with previous studies within uncertainties. The generally much higher portion of ordinary chondrite in Mars as compared to Earth underscores the differences in bulk composition between the two planets. Other work suggested that the highly reduced and volatile depleted enstatite chondrites were formed in the inner region of the solar system (closer than 1.5 AU), and more oxidised and volatile-rich ordinary chondrite sources originated in a region further away (probably $>$ 1.5 AU) \citep{morbidelli2012building,rubie2015accretion,fischer2017ruthenium}. Consequently, \cite{brasser2017cool} argued that the formation region of Mars is likely to be further from the Sun than Earth owing at least to the fact that its composition has a much higher computed fraction of ordinary chondrite.

\section{Framework}
\label{subsec:frame}
Assuming that Mars formed distantly from Earth’s feeding zone, we first investigate the possibility of producing differential Earth and Mars isotopic compositions with both the Classical and the Grand Tack model from the results of our $N$-body simulations. This is accomplished via comparison of the accretion zone of the Earth and Mars analogues generated in the simulation as described in the following section. Next, we estimate the average percentage of enstatite and ordinary chondrite incorporated into Earth and Mars by assuming the initial solid disk is composed of only enstatite in the inner region and ordinary chondrite in the outer region; alternatively, we test a scenario wherein the disk is effectively mixed with planetesimals and embryos of random compositions (see Sect. \ref{sec:chon_frac}). Finally, we calculate the expected isotopic composition of $^{17} \rm O$, $^{50} \rm Ti$, $^{54} \rm Cr$, $^{142} \rm Nd$, $^{64} \rm Ni$ and $^{92} \rm Mo$ in the martian mantle (Sect. \ref{sec:iso_mars}). Our purpose in this study is to examine how the orbital dynamics affects Mars' isotopic composition.

\section{Comparison of accretion zones between Earth and Mars}
\label{sec:accret_com}
\subsection{Simulation set up}
\label{subsec:simulation}

To estimate the possibility of reproducing Mars with a composition that is different from Earth’s, we analyse the data from a high number of $N$-body simulations for the Grand Tack and the Classical model as described above. The simulations and data used to examine the Grand Tack model were previously discussed in \cite{brasser2016analysis}, and we refer the reader to that work. We first briefly summarise their initial conditions and the organisation of the simulations.

In \cite{brasser2016analysis}, the formation of terrestrial planets were studied in the Grand Tack scenario by focusing on the traditional dynamic perspectives, which are the final orbital architecture and the masses of the terrestrial planets. The tack location for Jupiter at both 1.5 AU and 2 AU was considered, where ‘tack’ location is the position at which Jupiter reversed its migration. Two types of initial condition for planetestimals and embryos were adopted: the equal-mass initial condition, and the oligarchic initial condition (embryo resulting from the oligarchic growth of planetesimals). \cite{brasser2016analysis} employed the initial condition of embryos and planetesimals from \cite{jacobson2014lunar} for the equal-mass initial condition, in which embryos in the same simulation have the same initial masses. The initial embryo masses are either 0.025, 0.05 or 0.08 $M_{\Earth}$, where $M_{\Earth}$ is the mass of Earth. All the embryos within the disk were embedded in a disk of planetesimals. The total mass ratio of the embryos to planetesimals is either 1:1, 4:1, or 8:1. In total there are nine different subsets of initial conditions for the equal-mass initial condition. They performed 16 simulations in each subset; in total 144 simulations for each tack location. Table \ref{tb:embryo_no_1} shows the initial number of embryos in these nine subsets of initial conditions. The number of embryos increases with the mass ratio of embryos to planetesimals but decreases with the masses of embryos. They kept the number of planetesimals at 2000. The initial densities of embryos and planetesimals are 3 g $\rm cm^{-3}$.

\begin{table*}
%\tablewidth{0pt}
\caption{Initial number of embryos in each subset of the simulations with equal-mass embryos. They are classified by different masses of embryos, $M_{emb}$  and different mass ratio of embryos to planetesimals, $M_{emb}$:$M_{pl}$.}
\label{tb:embryo_no_1}
\centering
\begin{tabular}{lccc}
\hline\hline
 &  & $M_{emb}$:$M_{pl}$ & \\
 $M_{emb}$ & 1:1 & 4:1 & 8:1\\
\hline
0.025 $M_{\Earth}$ & 86 & 159 & 211 \\
0.05 $M_{\Earth}$  & 43 & 80  & 106 \\
0.08 $M_{\Earth}$  & 27 & 50  & 66  \\
\hline
\end{tabular}
\end{table*} 

In addition to forming terrestrial planets with equal-mass embryos, \cite{brasser2016analysis} also ran simulations with the initial conditions that are computed from the traditional oligarchic growth of the planetesimals \citep{kokubo1998oligarchic}. \cite{kokubo1998oligarchic} found that the orbital repulsion between embryos keeps their mutual separation wider than about 5 $R_{\rm Hill}$ during the oligarchic growth of the planetesimals, where $R_{\rm Hill}$ is the mutual Hill radii of two adjacent embryos. The oligarchic initial condition is more realistic compared to the equal mass initial condition because it is a result from the actual study of planetary embryo formation. \cite{brasser2016analysis} applied the semi-analytical oligarchic approach of \cite{chambers2006semi}. First, they calculated the total solid mass between 0.7 and 3 AU according to the minimum mass solar nebula $\Sigma_s$ = 7 g $\rm cm^{-2} \rm (a/1 AU)^{-3/2}$ \citep{hayashi1981structure}. Second, they increased the solid density at the ice line by a factor of three following the study of \cite{ogihara2009n}. The ice line was assumed to be static at 2.7 AU. Third, they set the spacing between the embryos to 10 $R_{\rm Hill}$ based on the results of \cite{kokubo1998oligarchic}; although in our work we ran additional runs with 7 and 5 $R_{\rm Hill}$. The spacing was computed assuming that embryos had their isolation masses, $m_{\rm iso} = 2\pi ab \Sigma_s$ , where $a$ is the semi-major axis of embryo and $b$ is the spacing between adjacent embryos \citep{chambers2006semi}. Here $m_{\rm iso}$ represents the largest mass of an embryo if it accretes all the solid mass within the annulus 2$\pi ab$. The semi-major axis of embryo $n$ is $a_n$ = $a_{n-1}[1 + b(2m_{\rm iso}/3M_{\odot})]$, which nearly follows a geometric sequence, where $M_{\odot}$ is the solar mass.

\begin{figure*}
\sidecaption
\includegraphics[width=12cm]{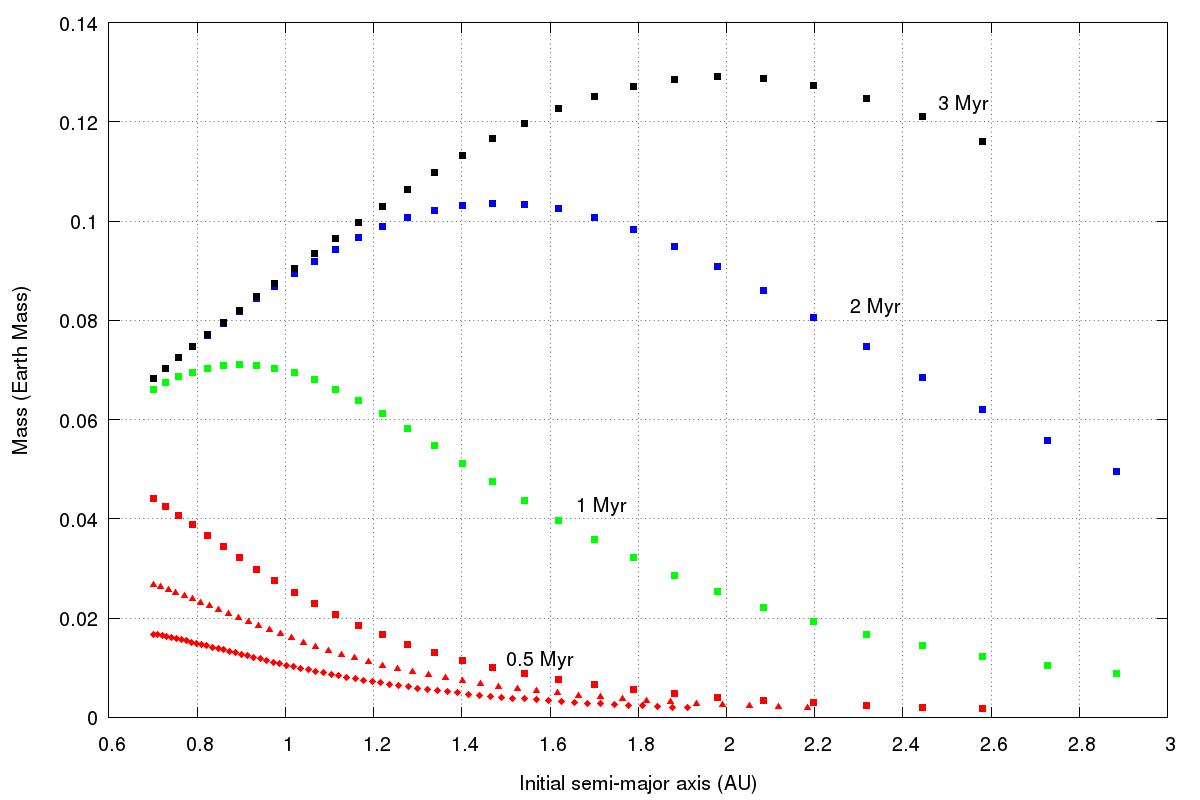}
\caption{The initial masses of embryos against their initial semi-major axes for the oligarchic initial condition. The red points are embryos in a 0.5 Myr disk, green points are in a 1 Myr disk, blue points are in a 2 Myr disk and black points are in a 3 Myr disk. Different point shapes represent different separation between adjacent embryos. Squares depict embryos separated by 10 $R_{\rm Hill}$ from each other, triangles are embryos with spacing of 7 $R_{\rm Hill}$ and rhombuses are embryos with spacing of 5 $R_{\rm Hill}$. The comparison of different spacing between embryos is only plotted for the 0.5 Myr disk.}
\label{fig:em_a_mass}
\end{figure*}

Other than calculating the spacing and the semi-major axis of embryos, the initial masses of embryos also needed to be computed. Following \cite{chambers2006semi}, the mass of embryos increases up to their isolation mass as 

\begin{equation}
m_{\rm p}(t) = m_{\rm iso}{\rm tanh^3} (t/\tau),
\label{eq:m_p(t)}  
\end{equation}
where $\tau$ is the growth timescale, which depends on the semi-major axis $a$ of the embryo, embryo spacing $b$, solid surface density $\Sigma_s$ and the radii of planetesimals that assembled into the embryo, and $t$ is the age of the solid disk before the start of the migration of Jupiter. \cite{chambers2006semi} assumed a planetesimal size of 10 km in calculating $\tau$ , which was also adopted by \cite{brasser2016analysis}. By substituting $t$ = 0.5 Myr, 1 Myr, 2 Myr or 3 Myr into Eq. (\ref{eq:m_p(t)}), they obtained four different subsets of initial conditions for the oligarchic growth scenario. They performed 16 simulations for each $t$, 64 in total for each tack location of Jupiter with 10 $R_{\rm Hill}$ embryo spacing. We perform more simulations with 5 and 7 $R_{\rm Hill}$ embryo spacing (64 each), but only for tack location of Jupiter at 1.5 AU. Figure \ref{fig:em_a_mass} shows the initial masses of the embryos as a function of their initial semi-major axes for the oligarchic initial condition. Different colors correspond to different ages of the solid disk. By comparing the squares with different colors, we observe that embryos with similar initial semi-major axes have larger masses if the solid disk is older. This is because embryos accrete more planetesimals within the disk if the solid disk exists for a longer time before the start of Jupiter’s migration. The initial number of embryos is larger in the 1 Myr and 2 Myr disk (see Table \ref{tb:embryo_no_2}). We also compare the initial number of embryos with different initial $R_{\rm Hill}$ separation in Fig. \ref{fig:em_a_mass}, but only for the 0.5 Myr disk. Comparing the red points with different shapes, the mass of embryos is lower but the initial number of embryos is larger if they have smaller mutual separation (see Table \ref{tb:embryo_no_2}).

\begin{table*}
%\tablewidth{0pt}
\caption{Initial number of embryos in each subset of the simulations with oligarchic embryos. They are classified by different ages of the solid disk before the migration of Jupiter and different spacing between adjacent embryos in units of mutual Hill radii, $R_{\rm Hill}$.} 
\label{tb:embryo_no_2}
\centering
\begin{tabular}{lccc}
\hline\hline
 &  & Mutual Hill radii spacing & \\
Disk age ($t$) & 10 $R_{\rm Hill}$ & 7 $R_{\rm Hill}$ & 5 $R_{\rm Hill}$\\
\hline
0.5 Myr & 29 & 43  & 63 \\
1 Myr   & 31 & 52  & 86 \\
2 Myr   & 31 & 52  & 86 \\
3 Myr   & 29 & 49  & 80 \\
\hline
\end{tabular}
\tablefoot{The total mass of solids is the same in each subset of simulation.}
\end{table*} 

The system consisting of gas giants, planetary embryos, and planetesimals was then simulated with the symplectic integrator package SyMBA \citep{duncan1998multiple} for 150 Myr using a time step of 7.3 days. The migration of Jupiter and Saturn was mimicked through the fictitious forces \citep{walsh2011low}. The gas disk developed by \cite{bitsch2015structure} was adopted, which is different from the one adopted by \cite{walsh2011low} and has a higher surface gas density. The initial gas surface density profile scales as $\Sigma(r)$ $\propto$ $r^{-1/2}$ and the temperature profile is $T(r)$ $\propto$  $r^{-6/7}$, where $r$ is the radial distance to the Sun. Both gas surface density and temperature decay until 5 Myr after the start of simulation, after which the disk is artificially photo-evaporated away with the $e$-folding time 100 kyr \citep{bitsch2014stellar}. The existing gas disk caused type-I migration and tidal damping of the eccentricities of the planetary embryos. The planetesimals are affected by the gas drag, for which \cite{brasser2016analysis} assumed each planetesimal had a radius of 50 km.

We also study the Classical model, but only for the oligarchic initial condition due to the limited computational resources. In other words, we used the same oligarchic initial conditions but placed Jupiter and Saturn on their current orbits. The time step and the total simulation time for each simulation in the Classical model is the same as the Grand Tack model. We test the Classical model with 10 $R_{\rm Hill}$ and 5 $R_{\rm Hill}$ mutual spacing between embryos.

\subsection{Definition of accretion zone}
\label{subsec:defin_accret}
We compare the composition of the planets that formed in the current region of the terrestrial planets. Assuming the disk follows a simple relation between isotopic ratios and distance to the Sun (see Sect. \ref{subsec:comp_diff}), the bulk composition of each planet can be roughly quantified. The compositional differences between planets are quantified by tracking the initial heliocentric distance of the material (either embryos or planetesimals) accreted into each planet. Therefore, the mass-weighted mean initial semi-major axis of incorporated material,

\begin{equation}
a_{\rm mwmi} = {\sum_i^N m_i a_i \over \sum_i^N m_i}  
\label{eq:a_mwmi}  
,\end{equation} 
was calculated for each terrestrial planet analogue, where $m_i$ and $a_i$ are the mass and initial semi-major axis of the accreted object $i$, and $N$ is the total number of embryos and planetesimals incorporated into the planet \citep{kaib2015feeding,brasser2017cool,fischer2018radial}. Planets with similar $a_{\rm mwmi}$ should have similar bulk composition. The weighted standard deviations of $a_{\rm mwmi}$,

\begin{equation}
\sigma_{\rm w}^2 = {{\sum_i^N m_i (a_i - a_{\rm mwmi})^2} \over {{(N-1) \over N} \sum_i^N m_i}}
\label{eq:sigma_w},
\end{equation}
are also calculated for each planet analogue. $a_{\rm mwmi}$ $\pm$ $\sigma_{\rm w}$ provides us the accretion zone of the planet analogue, which is the main region of the disk that the planet analogue sampled; $\sigma_{\rm w}$ is defined as zero if the planet did not accrete any planetesimal. 

\subsection{Results and discussion}
\label{subsec:results}
\subsubsection{Classical:Oligarchic initial condition}
\label{subsubsec:classical}
\begin{figure*}
\sidecaption
\includegraphics[width=12cm]{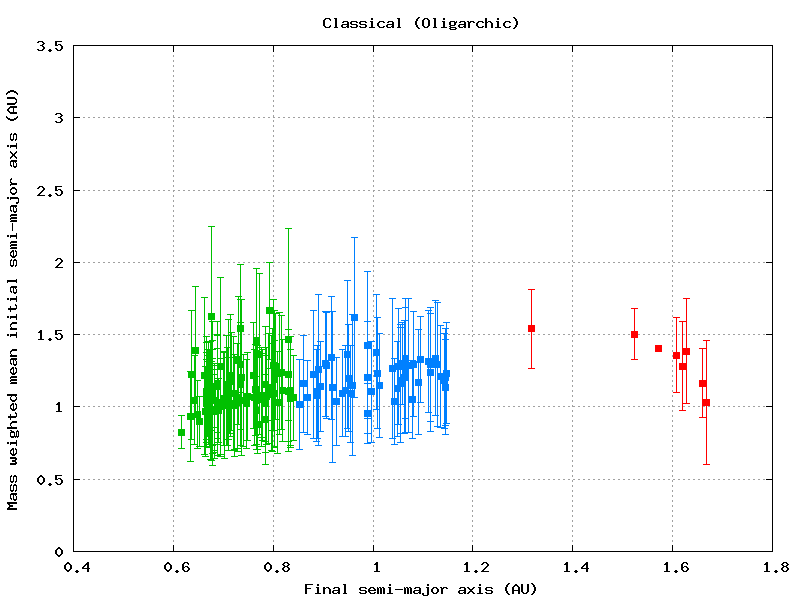}
\caption{Mass-weighted mean initial semi-major axis ($a_{\rm mwmi}$) of material accreted into each Venus analogue (green), Earth analogue (blue) and Mars analogue (red) as a function of their final semi-major axis for the oligarchic initial condition of the Classical model. We combine the results of 10 and 5 $R_{\rm Hill}$ and different ages of the solid disk (either $t$ = 0.5, 1, 2 or 3 Myr) within this plot.}
\label{fig:classical}

\end{figure*}

Figure \ref{fig:classical} depicts the $a_{\rm mwmi}$ of Venus analogues (green), Earth analogues (blue) and Mars analogues (red) as a function of their final semi-major axes for the oligarchic initial condition of the Classical model with Jupiter and Saturn on their current orbits. The results of initial embryos with spacing of 10 $R_{\rm Hill}$ and 5 $R_{\rm Hill}$ spacing, as well as different disk ages $t$ are combined here. Venus, Earth, and Mars analogues are defined as having masses and semi-major axes within the ranges (0.4 $M_{\Earth}$ $<$ $m_p$ $<$ 1.2 $M_{\Earth}$, 0.55 AU $<$ $a$ $<$ 0.85 AU), (0.5 $M_{\Earth}$ $<$ $m_p$ $<$ 1.5 $M_{\Earth}$, 0.85 AU $<$ $a$ $<$ 1.15 AU) and (0.05 $M_{\Earth}$ $<$ $m_p$ $<$ 0.15 $M_{\Earth}$, 1.3 AU $<$ $a$ $<$ 1.7 AU) \citep{brasser2016analysis}. The error bars are $\pm$ 1$\sigma_w$, which represent the accretion zone of each planet.

Comparing the data in Fig. \ref{fig:classical}, we find that most of the Earth (blue squares) and Venus (green squares) analogues have 1 AU $<$ $a_{\rm mwmi}$ $<$ 1.5 AU and their error bars share a similar range, which implies that the accretion zones, and therefore their bulk compositions, are likely to be similar. There is a weak trend of planets with larger final semi-major axes having larger $a_{\rm mwmi}$. A similar but stronger trend has been observed in the simulations performed by \cite{kaib2015feeding}. One possible reason is that our initial solid disk truncates at 0.7 AU, while the inner boundary of Kaib \& Cowan’s disk is at 0.5 AU. If the initial solid disk is extended to a region closer to the Sun, the Venus analogues should accrete more material originated from the inner-most region of the disk. Therefore, most of the Venus analogues would sample the disk closer to the Sun than the Earth analogues do and generate a more obvious linear trend between $a_{\rm mwmi}$ and the final semi-major axis. Comparing Earth (blue squares) with Mars (red squares) in Fig. \ref{fig:classical} leads us to a similar conclusion, since six out of eight of the Mars analogues have 1 AU $<$ $a_{\rm mwmi}$ $<$ 1.5 AU and the error bars of all the red and blue squares share a similar range. Therefore, the Classical model predicts a similar bulk composition between Venus, Earth, and Mars. That is to say, the Classical model inadequately explains the bulk composition difference between Earth and Mars if we are able to form Mars with its current mass. If this restriction is ignored, it is possible to form a distinct Earth and Mars (See Sect. \ref{sec:chon_frac}). 

Table \ref{tb:planet_stat} lists the statistics for all sets of simulations performed in this study. The number of Venus and Earth analogues we obtained from the Classical model is similar to the oligarchic simulation of the Grand Tack model. Although 102 planets are formed in the orbital parameter space we set for Mars analogues, only 8 fall within a satisfactory mass range for Mars within the 128 Classical's simulations performed. This is far fewer than those obtained from the Grand Tack simulations. The reason is that most of the planets formed near Mars’ current position have masses larger than the value that we defined for Mars analogues ($>$ 0.15 $M_{\Earth}$). The average mass of the planets within 1.3 $<$ $a$ $<$ 1.7 AU in the simulations is 0.45 $M_{\Earth}$, which is more than three times the mass of Mars. As previously mentioned, the biggest problem of the Classical model is forming a Mars that is too massive \citep[e.g.][]{chambers2001making,raymond2009building}.

\begin{table*}

%\tablewidth{0pt}
\caption{The total number of planets and terrestrial planet analogues formed in each set of simulations.}
\label{tb:planet_stat}
\begin{tabular}{lcccccc}

\hline\hline
Simulation set & No. of & Planets & Venus & Earth & Mars \\
 & simulations & & analogues & analogues & analogues\\
\hline
Grand Tack &  &  &  &  & \\
Equal mass embryos & 144 & 635 & 151 & 62 & 59\\
Tack at 1.5 AU &  &  &  &  & \\ \hline
Grand Tack &   &  &  &  & \\
Equal mass embryos & 144 & 685 & 121 & 90 & 32 \\
Tack at 2 AU &  & & & & \\ \hline
Grand Tack &  &  &  & \\
Oligarchic 10 $R_{\rm Hill}$ & 64 & 210 & 70 & 22 & 17 \\
Tack at 1.5 AU &  &  &  & \\ \hline
Grand Tack &  &  &  & \\
Oligarchic 10 $R_{\rm Hill}$ & 64 & 273 & 57 & 37 & 14 \\
Tack at 2 AU  &  &  & & \\ \hline
Grand Tack   &  &  &  &  \\
Oligarchic 7 $R_{\rm Hill}$ & 64 & 231 & 59 & 27 & 18 \\
Tack at 1.5 AU &  &  &  &  \\ \hline
Grand Tack   &  &  &  &  \\
Oligarchic 5 $R_{\rm Hill}$ & 64 & 250 & 59 & 22 & 19 \\
Tack at 1.5 AU  &  &  &  &  \\ \hline
Classical   \\
Oligarchic 10 $R_{\rm Hill}$ & 64 & 225 & 50 & 22 & 4 \\ \hline
Classical \\ 
Oligarchic 5 $R_{\rm Hill}$ & 64 & 253 & 45 & 30 & 4  \\ \hline
\end{tabular}
\tablefoot{Please refer to Sect. \ref{subsec:simulation} for the classification of the simulation groups.}
\end{table*}

\subsubsection{Grand Tack: Oligarchic initial conditions}
\label{subsubsec:grand_oli}
\begin{figure*}
\sidecaption
\includegraphics[width=12cm]{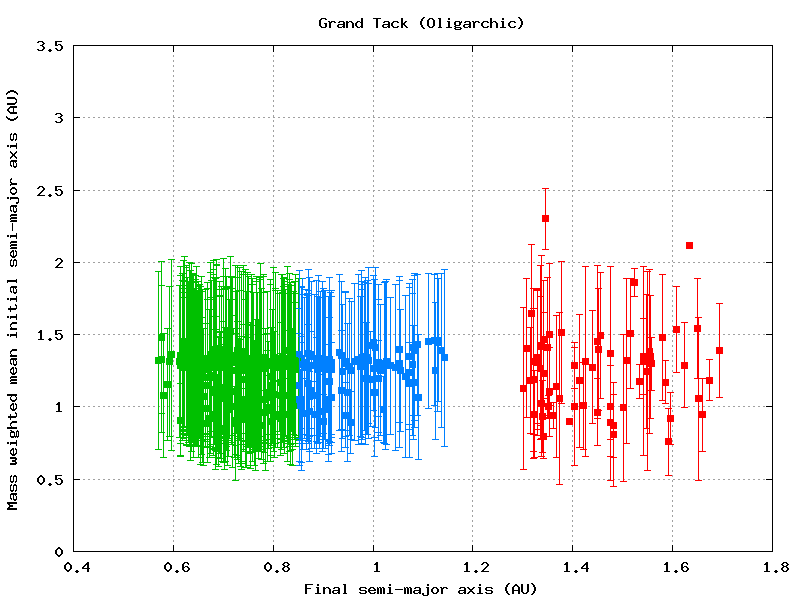}
\caption{As in Fig. \ref{fig:classical}, but for the Grand Tack model with oligarchic initial conditions. The results of Jupiter tacked at 1.5 AU or 2 AU, as well as different initial spacing between adjacent embryos (either 10, 7 or 5 $R_{\rm Hill}$) and different ages of the solid disk (either $t$ = 0.5, 1, 2 or 3 Myr) are combined.}
\label{fig:grand_oli}

\end{figure*}

Figure \ref{fig:grand_oli} resembles Fig. \ref{fig:classical}, but depicts the results from the Grand Tack simulations with oligarchic initial conditions. The results of the tack at 1.5 AU and 2 AU of Jupiter, as well as different adjacent embryo spacing and disk ages, are all plotted in Fig. \ref{fig:grand_oli}. Similar to Fig. \ref{fig:classical}, most of the Venus and Earth analogues have $a_{\rm mwmi}$ between 1 and 1.5 AU. A few of them have $a_{\rm mwmi}$ smaller than 1 AU; these sample the innermost portion of the disk. Statistics of the Mars analogues suggest that their bulk composition is likely to be similar to both Earth and Venus since only $\sim$13\% of the Mars analogues within the oligarchic simulations have $a_{\rm mwmi}$ $>$ 1.5 AU, and their error bars in Fig. \ref{fig:grand_oli} mostly share the same range with the Venus and the Earth analogues. Only two Mars analogues in our output have $a_{\rm mwmi}$ $>$ 2 AU. We return to this type of Mars analogue below (Sect. \ref{subsubsec:grand_equal}). According to simulations, the bulk compositions of these three planets are likely to be similar.

It is worth pointing out that about one-fourth of the Mars analogues have $a_{\rm mwmi}$ $\leq$ 1 AU. The fraction is even higher than the fraction of the Earth analogues with $a_{\rm mwmi}$ $\leq$ 1 AU. Therefore, the oligarchic simulation results show that it is possible for Mars to be formed in the innermost region of the protoplanetary disk and then scattered to its current location. This was also suggested by \cite{hansen2009formation} and \cite{fitoussi2016building} and might be taken as a natural explanation for the formation of Mars, that is if we do not consider its isotopic compositional difference from Earth \citep{brasser2017cool}.

%The results from both tack scenarios at 1.5 AU and at 2 AU for Jupiter would seem to suggest that Mars’ bulk composition should be similar to Earth’s. Still, the statistics from these two sets of simulations are slightly different. \textbf{Simulations for tack at 2 AU generate at least 50 more planets than tack at 1.5 AU for the oligarchic and the equal mass initial condition, respectively.} This is because Jupiter migrated into an inner region if it tacked at 1.5 AU and thus scattered more planetesimals or embryos out to a distance beyond 5 AU. Thus, less solid mass is left in the inner disk for planet formation. We find, however, that simulations with a tack at 1.5 AU yield more Mars and Venus analogues than the tack occurred at 2 AU. It is probably caused by Jupiter clearing the disk in between 1 to 1.5 AU more effectively and pushing more solid material within 1 AU by mean-motion resonances and scattering if the tack location is at 1.5 AU. Consequently, the Mars analogues would not be too massive and the Venus analogues can accrete enough mass if the tack location is at 1.5 AU.

Reducing the spacing of embryos (from 10 $R_{\rm Hill}$ to either 7 $R_{\rm Hill}$ or 5 $R_{\rm Hill}$) produces different sets of initial conditions with lower average individual embryo mass and a higher initial number of embryos (See Fig. \ref{fig:em_a_mass} and Table \ref{tb:embryo_no_2}). Table \ref{tb:planet_stat} shows the statistics of 5 and 7 $R_{\rm Hill}$ spacing for the oligarchic Grand Tack simulations. The statistics of both 5 and 7 $R_{\rm Hill}$ spacing are similar to that of 10 $R_{\rm Hill}$, which indicates that the final results are not strongly dependent on the initial conditions of the embryos.

\subsubsection{Grand Tack: Equal mass initial condition}
\label{subsubsec:grand_equal}
\begin{figure*}
\sidecaption
\includegraphics[width=12cm]{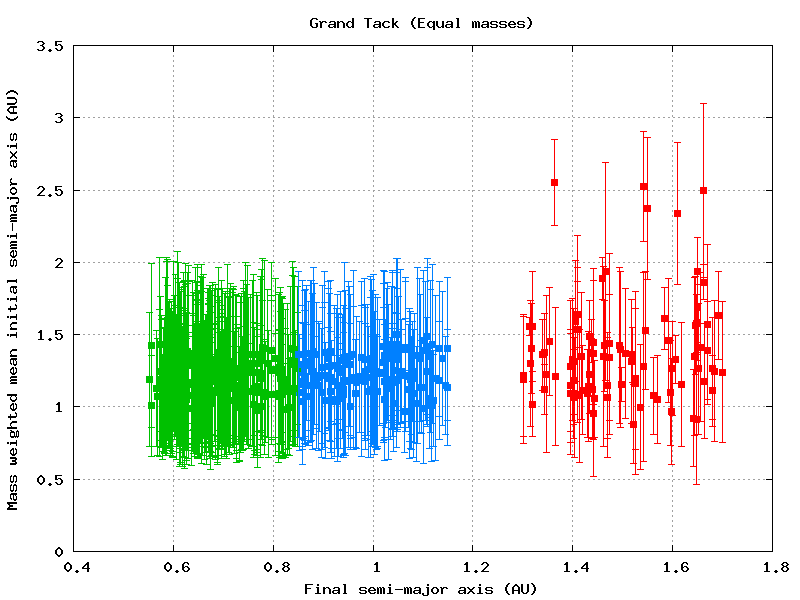}
\caption{As in Fig. \ref{fig:grand_oli} but for the equal-mass initial conditions. This plot is similar to Figs. 2 and 8 in \cite{brasser2017cool}, but we only show the Venus (green), Earth (blue) and Mars (red) analogues.}

\label{fig:equal_mass}

\end{figure*}

Figure \ref{fig:equal_mass} is similar to Fig. \ref{fig:grand_oli}, but shows the result of using the equal-mass initial conditions of the Grand Tack model where Jupiter tacked at either 1.5 AU or 2 AU. This plot resembles the previously published Figs. 2 and 8 of \cite{brasser2017cool}, but we highlight the terrestrial planet analogues in our plot.

Similar to the oligarchic initial condition, we find that nearly all of our Venus, Earth, and Mars analogues have error bars that share a similar range of uncertainty. As such, most of the terrestrial planet analogues can be seen to have sampled a similar region of the disk. The value $a_{\rm mwmi}$ needs to be $>$ 2 AU if the error bars of the Mars analogues are not in the same range as all of the Venus and Earth analogues. The Mars analogues with $a_{\rm mwmi}$ $>$ 2 AU are the ones that we are searching for because those Mars analogues would have a very different accretion zone with the Venus and Earth analogues. They formed in the outer region of the disk and accreted material mainly from there. Then Jupiter scattered them close to the current location of Mars when it migrated outward. This did not occur for Earth since we do not find any Earth analogue with $a_{\rm mwmi}$ $>$ 2 AU. There are seven of the Mars analogues with $a_{\rm mwmi}$ $>$ 2 AU in total in the Grand Tack simulations (including all the simulations with equal mass and oligarchic initial condition). We find none of them in the Classical simulations.

\begin{figure*}

\resizebox{\hsize}{!}{\includegraphics[width=1.1\textwidth]{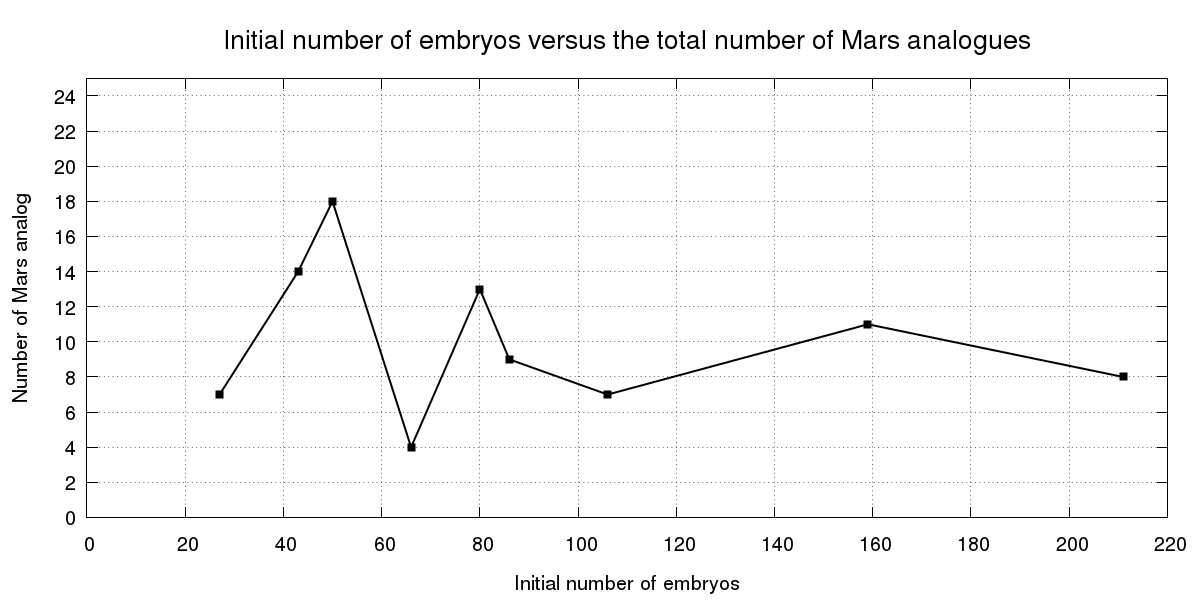}}
\caption{The total number of Mars analogues in each subset of simulations as a function of the initial number of embryos per simulation (see Table \ref{tb:embryo_no_1}) of the Grand Tack model with equal-mass initial conditions. The results from simulations with Jupiter’s tack location at 1.5 AU and 2 AU are combined.}    

\label{fig:embryos_mars_analog}

\end{figure*}

We obtained 635 planets in total when the tack location was at 1.5 AU and 685 planets when the tack location was at 2 AU, which greatly exceed the results from the oligarchic initial conditions (See Table \ref{tb:planet_stat}). The main reason for this is that we have more simulations per set when using the equal-mass initial conditions. We only performed 64 simulations per set for the oligarchic initial conditions, whereas we had 144 simulations for the equal-mass initial conditions. The oligarchic initial conditions, however, also yield fewer planets per simulation. This is because the initial number of embryos in the oligarchic case is usually fewer than the equal-mass case (comparing the numbers between Table \ref{tb:embryo_no_1} and \ref{tb:embryo_no_2}). We would expect that putting more embryos at the beginning of the simulation would help to form more planets at the end. Yet, this is not the case when we consider the formation of the Mars analogues. Instead, we find that increasing the number of embryos does not help to produce more Mars analogues. Figure \ref{fig:embryos_mars_analog} shows the total number of Mars analogues in each subset of simulations against the initial number of embryos per simulation (see Table \ref{tb:embryo_no_1} for the initial number of embryo adopted for each subset with equal-mass embryos). We have 9 subsets and therefore 9 different initial numbers of embryos: from 27 to 211 embryos. Each subset consists of 32 simulations, in which 16 are simulations with Jupiter’s tack location at 1.5 AU and the other 16 are simulations with Jupiter’s tack location at 2 AU. Even if we start the simulations with more than 200 embryos, the total number of Mars analogues generated were fewer than the simulations with only about 50 embryos. Therefore, there is no correlation between the initial number of embryos and the final number of Mars analogues, which violates our earlier prediction that more embryos at the beginning of simulations would help to form more Mars analogues.

\subsection{Summary}
\label{subsec:sum_accret_com}
Here we examined the possibility of forming Earth and Mars with different accretion zones and therefore different bulk compositions in the Classical and the Grand Tack models. We find that simulations in the Classical model fail to form any of the Mars analogues with an accretion zone distinct from the Earth analogues ($a_{\rm mwmi}$ $>$ 2 AU) when we only consider Mars analogues close to its current mass. Therefore, the Classical model has trouble reproducing the compositional difference between Earth and Mars. It may not be valid, however, to make such a conclusion since we only have eight Mars analogues in the Classical model due to fewer numbers of simulations and its own drawback in explaining the small mass of Mars. Therefore, we carry out another study comparing the bulk composition of Earth and Mars, presented in the following section (see Sect. \ref{sec:chon_frac}). The Grand Tack model has a slightly higher success rate in producing compositionally different Earth and Mars because we obtain seven Mars analogues with accretion zones that are wholly distinct from that of the Earth analogues, although statistically the Grand Tack model has a much higher probability to form Earth and Mars with a similar accretion zone. Both the Classical and the Grand Tack models have difficulty forming Earth and Mars with quite different accretion zones that then lead to different bulk compositions. 

\section{Chondrite fractions to make Earth and Mars}
\label{sec:chon_frac}
%\subsection{Homogeneous solid disk}
%\label{sec:homo_disk}

We now calculate the bulk composition of Earth and Mars from the output of our dynamical simulations. We assume that the solid disk was initially composed of only enstatite chondrite and ordinary chondrite. 
%The major reason for this assumption is that the isotopic data for the chondrite groups is the most abundant among different types of meteorite. Furthermore, the chondrites are generally accepted to represent leftover building blocks of planet formation, and as such are expected to represent a reliable sample of the solid masses in the early solar system. 
As mentioned in Sect. \ref{subsec:comp_diff}, it has been suggested -- but not universally accepted -- that the solid disk originally had a heliocentric composition gradient consisting of dry, reduced enstatite chondrite close to the Sun, moderately volatile-rich and oxidised ordinary chondrite in the region of the asteroid belt, and highly oxidised and volatile-rich carbonaceous chondrites beyond \citep{morbidelli2012building,rubie2015accretion,fischer2017ruthenium}, mirroring the heliocentric distribution of asteroid groups \citep{gradie1982compositional,demeo2014solar}. Due to their different water fractions, a plausible reason for the transition from enstatite chondrite to ordinary chondrite is snow line migration. However, if this were true then we would also expect the ordinary chondrites to have as much water as carbonaceous chondrites. This is not the case. In truth, we are not aware of any physical explanation for this proposed heliocentric compositional gradient, and we only make use of previously published results. 
%This may imply that the composition of the disk is homogeneous locally when the planetesimals formed. 

Although recent studies provide an estimate for the initial location boundary between these two categories of chondrite \citep{morbidelli2012building,fischer2017ruthenium,o2018delivery}, there is still debate on the exact location for the disk to change its composition from enstatite to ordinary. We define this transition's location as the break location of the disk. Asteroid 21 Lutetia, which is considered to be a candidate source of the enstatite chondrites \citep{vernazza2009plausible,coradini2011surface} is currently located at $\sim$2.4 AU. However, it is possible -- and consistent with our Grand Tack simulations -- that it formed in the inner solar-system and was subsequently scattered by either emerging protoplanets or by the migration of Jupiter \citep{vernazza2011asteroid}. Therefore, the break location of the disk may be closer to the Sun.

We therefore set this break location as a free parameter and assume that it ranges from 1 to 2 AU, with 0.1 AU intervals (i.e. if the break location of the disk is at 1.5 AU, all the planetesimals and embryos initially located within 1.5 AU initially are made of enstatite chondrite, and those initially located further than 1.5 AU are made of ordinary chondrite). We then make use of the simulation results in Sect. \ref{sec:accret_com} to calculate the average percentage of enstatite and ordinary chondrite incorporated into the Earth and Mars analogues for each break location.

\begin{figure*}

\resizebox{\hsize}{!}{\includegraphics{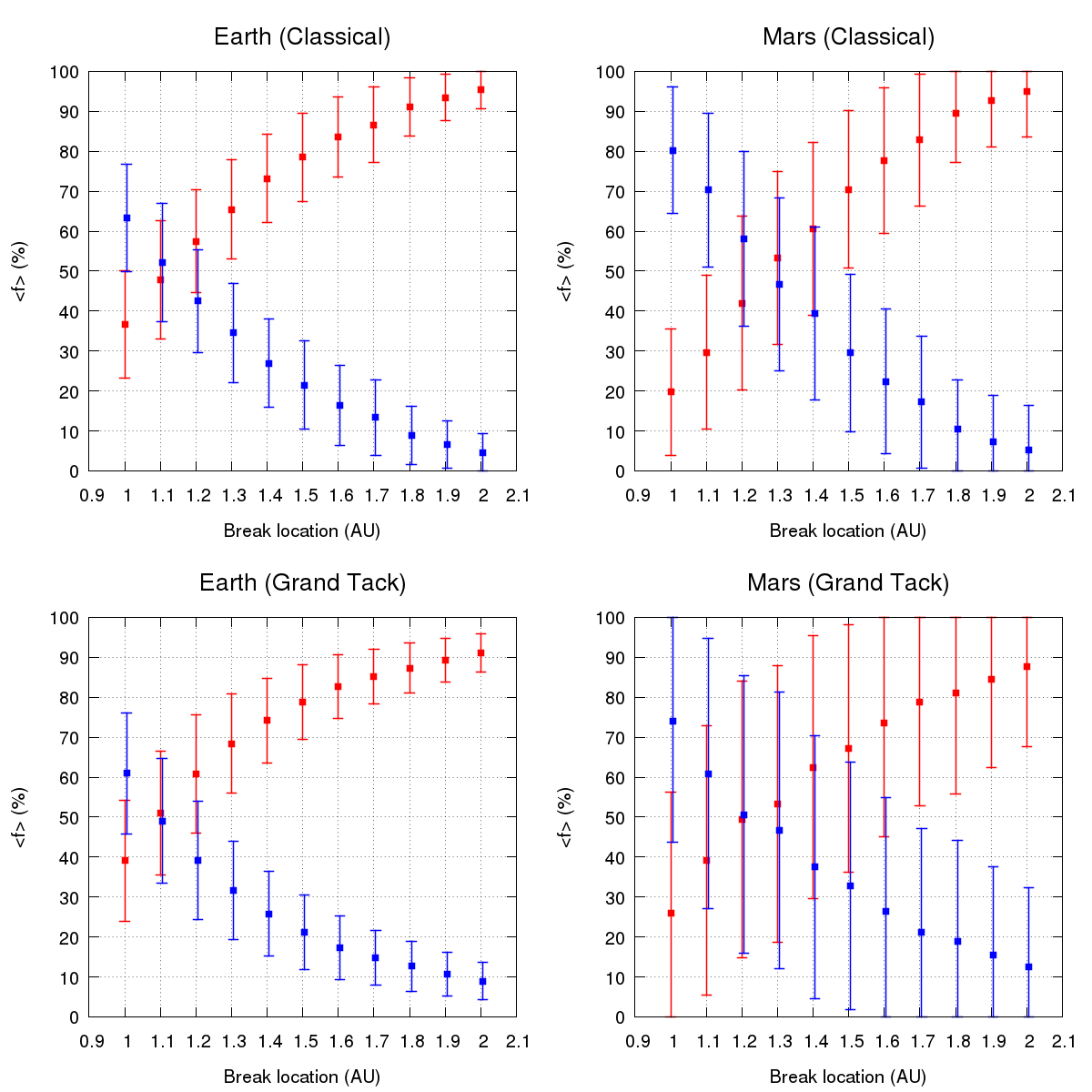}}
\caption{The average percentage contribution of enstatite chondrite (red) and ordinary chondrite (blue) to the bulk compositions of Earth and Mars. Data are plotted as a function of the distance where the initial solid disk changes its composition from enstatite chondrite to ordinary chondrite, which is defined as the break location of the disk. The error bars are $\pm$1$\sigma$. The two upper panels depict the average composition of Earth and Mars in the Classical model simulations. The two lower panels show the results from Grand Tack simulations with equal-mass initial conditions, which is similar to Fig. 7 of \cite{brasser2017cool}.}    

\label{fig:break_chon}

\end{figure*}

Figure \ref{fig:break_chon} shows the average percentage of enstatite chondrite (red) and ordinary chondrite (blue) incorporated into the Earth and Mars analogues with different break locations of the disk for both models. We only examine data from the simulations with the equal-mass initial condition for the Grand Tack model because the oligarchic initial condition provides similar results in the accretion zone analysis (see Sect. \ref{sec:accret_com}). We relax the constraints on the Mars analogues by not considering their masses for the Classical model in this study to ensure that enough data are generated. The total percentage of enstatite and ordinary chondrite at each break location sums up to 100$\%$. With a more distant break location, both Earth and Mars consist mainly of enstatite chondrite since the disk consists of more enstatite chondrite. The break location, however, cannot exceed 1.8 AU, otherwise the bulk composition of both Earth and Mars would consist of about 90$\%$ enstatite chondrite, which would make them too similar to each other and would also be inconsistent with Mars’ suggested bulk composition \citep{sanloup1999simple}. Therefore, the key here is whether or not a particular break location exists at which we can reproduce the measured isotopic composition for both Earth and Mars simultaneously. \cite{brasser2018GRL} reported a best-fit composition of $68\%^{+0}_{-39}$ enstatite chondrite plus $32\%^{+35}_{-0}$  ordinary chondrite for Mars, whereas Mars is suggested to be 45\% enstatite chondrite and 55\% ordinary chondrite according to \cite{sanloup1999simple} and \cite{tang201460fe}. The break location in between 1.1 to 1.5 AU for both models can satisfy the average bulk composition of Mars in Fig. \ref{fig:break_chon}. Combining this with the most recent results for Earth ($\sim$70\% enstatite chondrite and $\sim$25\% ordinary chondrite, \cite{dauphas2017isotopic}) leads us to conclude that the break location of the disk should be at 1.3 AU to 1.4 AU in both the Classical and the Grand Tack models. 

\begin{figure*}
\sidecaption
\includegraphics[width=12cm]{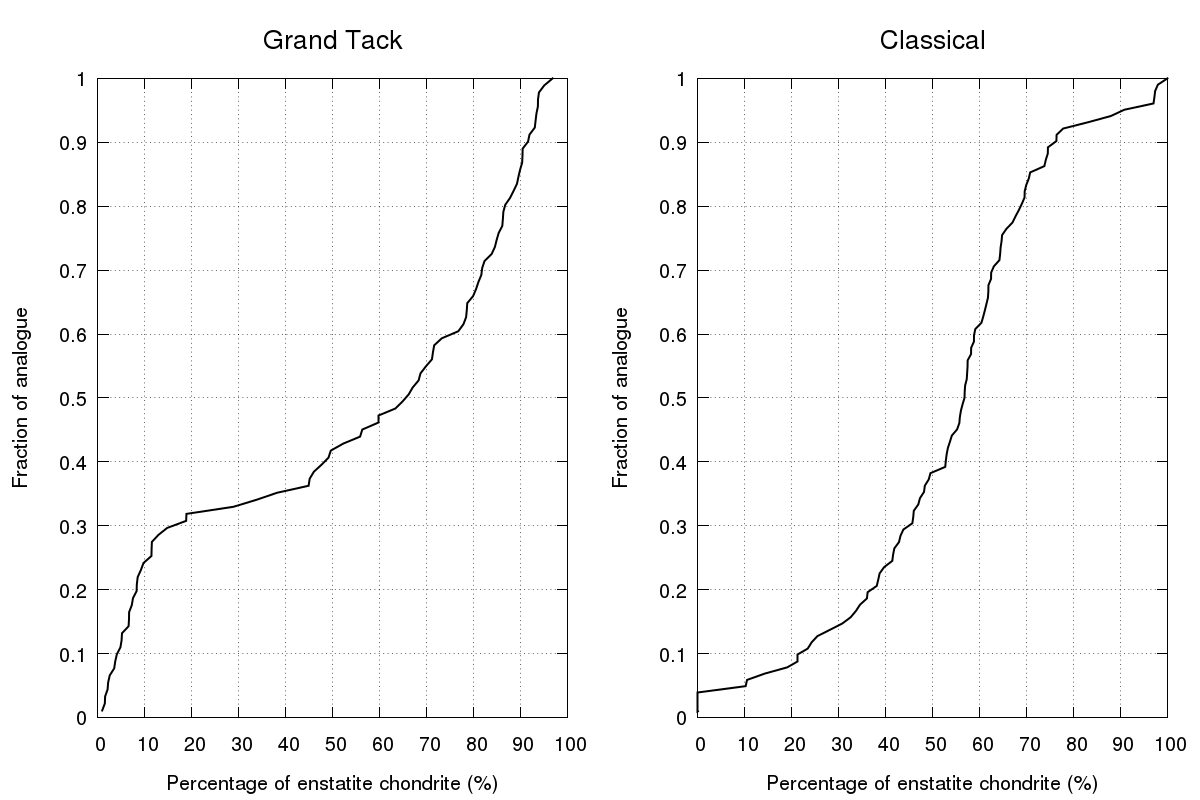}
\caption{The cumulative distribution function of the Mars analogues’ final composition (percentage of enstatite chondrite incorporated into each planet) when the break location of the disk is 1.3 AU. The left panel shows results from the Grand Tack simulations with initial equal-mass embryos and the right panel is results from the Classical simulations with initial oligarchic embryos.}     

\label{fig:cdf_mars}

\end{figure*}

It is important to note, however, that even if the \textit{average} mean composition of Mars analogues is close to 50:50 when the break location is at 1.3 AU, this result does not simply imply that most of the Mars analogues have such a composition. The error bars of Mars in Figure \ref{fig:break_chon} represent deviations of about $\pm$20\% to 30\% from the mean values, which suggest that we are dealing with averaging over two extreme cases that happen to meet in the middle. To understand the origin of the large scatter, we plot the cumulative distribution functions (CDFs) of the fraction of enstatite chondrite in the Mars analogues and determine the fraction of them that fall close to the 50:50 mark. Figure \ref{fig:cdf_mars} shows the result of this analysis, where the break location is at 1.3 AU. The left panel shows Mars analogues from the Grand Tack model with equal-mass initial conditions, and the right panel plots the CDF for Mars analogues from the Classical model with the oligarchic initial conditions. According to Fig. \ref{fig:break_chon}, the \textit{average} mean composition of Mars in both models is close to 50:50. In the Grand Tack, however, only $\sim$20\% of them fall in the range wherein they possess 29\% to 68\% enstatite chondrite by mass \citep{brasser2018GRL}, whereas $\sim$65\% of them do so in the Classical model. Therefore, the Grand Tack model does not work better than the Classical model in producing Mars’ documented composition if we neglect the massive Mars problem of the Classical model.

\begin{figure*}

\resizebox{\hsize}{!}{\includegraphics{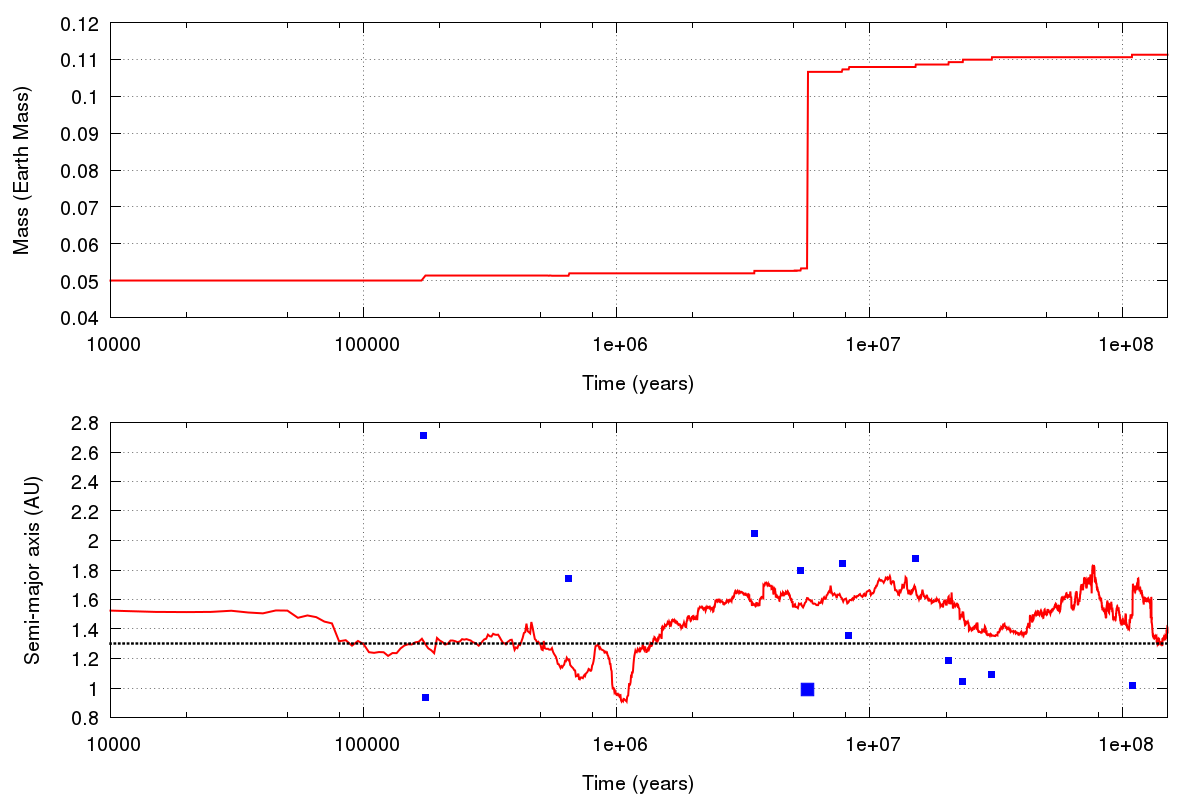}}
\caption{The evolution of mass and semi-major axis, respectively, of a Mars analogue (red solid lines in both panels) with a final 50:50 composition in the Grand Tack simulation with initial equal mass embryos. The squares in the lower panel represent the initial semi-major axes and the collision time with the Mars analogue of the planetesimals (smaller squares) and embryos (larger square). The black dotted horizontal line at 1.3 AU in the lower panel is the break location of the initial disk.}   

\label{fig:example}

\end{figure*}

To better understand the origin of Mars’ roughly 50:50 composition, we tracked its accretion history. By randomly picking 6 Mars analogues from the Grand Tack simulations and 5 from the Classical simulations, we find that in order to reproduce Mars’ suggested composition, 9 out of 11 involve collision between two embryos of different initial composition. Figure \ref{fig:example} shows an example of the mass and semi-major axis evolution of a Mars analogue from the Grand Tack simulation. The squares represent the initial semi-major axis of the planetesimals and embryos and the time they collided with the Mars analogue. Their sizes are scaled by their masses. In this case, the Mars analogue starts at $\sim$1.5 AU and is therefore an embryo composed of ordinary chondrite because the break location is at 1.3 AU. It stays near its initial location throughout its whole evolution. Its mass contributed by planetesimals made of enstatite chondrite is comparable to those made of ordinary chondrite: a total of seven of the planetesimals originated from the region further than the break location (squares above the horizontal dotted line at 1.3 AU) and five of the planetesimals started in the inner disk (squares below the horizontal dotted line at 1.3 AU), but they are not enough to increase its mass to the current value. Instead, the major reason for its 50:50 composition is a collision in the simulation at 5.7 Myr with another embryo whose composition is enstatite chondrite. This collision, doubling the mass of the Mars analogue is represented by a large square at $\sim$1 AU in the lower panel of Fig. \ref{fig:example}. This particular case agrees with the prediction by \cite{brasser2017cool} that Mars formed in a distant region and is therefore initially mainly composed of ordinary chondrite. As pointed out in that work, it was then scattered into the inner disk by Jupiter’s migration and possibly collided with an embryo mostly made of enstatite chondrite from the inner disk. This scenario naturally explains the isotopic compositional difference between Earth and Mars. Although collisions in $N$-body simulations occur at random, a colossal impact at 5.7 Myr is entirely consistent with Mars’ formation timescale.

\section{Using dynamics to produce the isotopic composition of Mars}
\label{sec:iso_mars}

Although the Grand Tack model is not a better model to explain the composition of Mars, its success in explaining the mass-distance distribution of the inner solar system should not be overlooked. Besides, if Mars formed in a distant region, it had to rely on the migration of Jupiter to scatter to its current location. 
%The next question we ask is whether the Grand Tack model can reproduce the isotopic composition of Mars’ mantle, which can be measured from the martian meteorites. The isotopic differences between Mars’ mantle and that of the Earth were measured for several elements, including $^{17} \rm O$, $^{50} \rm Ti$, $^{54} \rm Cr$, $^{64} \rm Ni$ and $^{92} \rm Mo$ \citep{dauphas2017isotopic,brasser2018GRL}. 

In the previous section, we computed the bulk composition of Mars from a mixture of both enstatite and ordinary chondrite. Even though this mixture agrees with published results of \cite{brasser2018GRL}, such compositions are extrapolated from the measurements of isotopic anomalies in Mars' mantle. Therefore a comparison between our computed composition of Mars from the dynamical simulations with available martian isotopic anomalies is warranted. Here we are interested not only in the predicted average values of the isotopic composition of Mars that result from the dynamical simulations, but also their uncertainties, and, furthermore, whether these uncertainties are comparable to the measured uncertainties, which may allow for additional insight regarding Mars' origin. Therefore, we proceed to calculate Mars' expected isotopic composition as follows. 

The isotopic anomaly is here defined as the deviation of the isotopic composition of a nuclide with respect to the terrestrial standard value. To obtain the correct isotopic composition for both Earth and Mars, the break location of the disk should be around 1.3 AU (see Fig. \ref{fig:break_chon}). We did not include carbonaceous chondrite in Sect. \ref{sec:chon_frac} because it only takes up a tiny portion of Earth and Mars according to a few isotopic studies \citep{warren2011stable,dauphas2017isotopic,dauphas2014calcium,sanloup1999simple}. When calculating the isotopic anomaly, however, the tiny contribution of carbonaceous chondrite may be enough in some cases to cause significant alteration in the value. To address this, we re-estimate the composition of Mars' analogues in this section by assuming the solids further than 3 AU are made of carbonaceous chondrite (CO and CV). We multiply the mass of each planetesimal further than 3 AU by ten since we have a relatively low-mass outer disk in the simulations. The isotopic composition of an element in Mars’ mantle $\varepsilon_{\rm Mars}$ can be calculated by
\begin{equation}
\varepsilon_{\rm Mars} = {\sum_i f_i c_i \varepsilon_i \over \sum_i c_i \varepsilon_i},
\label{eq:epsilon}  
\end{equation} 
where $f_i$ is the percentage of chondrite type $i$ in Mars ($i$ is either enstatite, ordinary, or carbonaceous chondrite), $c_i$ is the concentration of the element in chondrite type $i$ and $\varepsilon_i$ is the isotopic composition of the element in chondrite type $i$ \citep{dauphas2017isotopic}. The mean composition ($\pm$1$\sigma$) of Mars is $53\%^{+34}_{-34}$ enstatite chondrite, $47\%^{+35}_{-35}$ ordinary chondrite and $0.2\%^{+4.7}_{-0.2}$ carbonaceous chondrite if the break location of the disk is at 1.3 AU (see Fig. \ref{fig:break_chon}). The uncertainties in the final composition are the result of the great potential for variety in Mars' dynamical history, and in the $\varepsilon_i$ for each chondritic source. Therefore, a Monte Carlo method is adopted to calculate $\varepsilon_{\rm Mars}$ for each element. We pick the $f_i$ from our dynamcial simulations and $\varepsilon_i$ from measured values listed in \cite{dauphas2017isotopic}. We use the Box-Mueller transform to approximate a standard normal distribution and then calculate the corresponding $\varepsilon_{\rm Mars}$ from this distribution. We repeat this 60000 times and obtain a list of values for $\varepsilon_{\rm Mars}$ of an element. The mean and the standard deviation of $\varepsilon_{\rm Mars}$ of each elements is then calculated with this list of values. Our work is different from \cite{brasser2018GRL} since we adopt the composition of Mars from our dynamical simulation, whereas the results of \cite{brasser2018GRL} are based on a Monte Carlo Mixing model. 

Table \ref{tb:isotope} shows the average $\varepsilon_{\rm Mars}$ and the 2$\sigma$ uncertainty of each element. We compare the values that we calculated and those measured in different studies and summarised by \cite{dauphas2017isotopic}. The measured value for $\mu^{142} \rm Nd$ is the mean value of enriched shergottites in \cite{kruijer2017early}. We adopt the same computed value as \cite{brasser2018GRL} for $\varepsilon^{64}\rm Ni$, where they rely on existing $\varepsilon^{62}\rm Ni$ isotope data for Mars and the correlation between $\varepsilon^{64}\rm Ni$ and $\varepsilon^{62}\rm Ni$ \citep{tang201460fe}. The $\varepsilon^{64}\rm Ni$ of the martian mantle has not been measured precisely. We observe that all the elements fit within the 2$\sigma$ uncertainties of the measured values. For most elements, the computed isotopic uncertainties resulting from dynamics are comparable to the measured values. Oxygen, Titanium, and Neodymium are lithophile elements and therefore would remain in the mantle since they were accreted by Earth or Mars \citep{dauphas2017isotopic}. Chromium is moderately siderophile \citep[e.g.][] {righter2011moderately}. A portion of accreted Cr in the mantle sank into the core during the formation of Mars. This may alter the $\varepsilon_{\rm Mars}$ that we calculated from the full accretion of Mars. Nevertheless, the portion of Cr that sank into the core during Earth’s formation is tiny \citep{dauphas2017isotopic}. For simplicity, we treat Cr behaviour in Mars in a similar manner to that in Earth.

\begin{table*}
\centering
%\tablewidth{0pt}
%\tabletypesize{\scriptsize}
\caption{Comparison between the calculated and the average measured isotopic composition ($\varepsilon_{\rm Mars}$) from \cite{dauphas2017isotopic} for $\Delta^{17} \rm O$, $\varepsilon^{50} \rm Ti$, $\varepsilon^{54} \rm Cr$ and $\varepsilon^{92} \rm Mo$, \cite{kruijer2017early} for $\mu^{142} \rm Nd$ and \cite{brasser2018GRL} for $\varepsilon^{64} \rm Ni$ of Mars’ mantle. The isotopic anomaly of these elements is calculated from a break location of the disk at 1.3 AU in the Grand Tack model. Carbonaceous chondrite is also included in the calculation (see Sect. \ref{sec:iso_mars} for details). The uncertainties of the calculated values and the average measured values are 2$\sigma$.}
\label{tb:isotope}
\begin{tabular}{lcccccc}

\hline\hline\\
 & $\Delta^{17}\rm O$ & $\pm$ & $\varepsilon^{50}\rm Ti$ & $\pm$ & $\varepsilon^{54}\rm Cr$ & $\pm$ \\
\hline\\ 
Calculated & 0.43 & 0.33 & -0.37 & 0.21 & -0.14 & 0.13\\ 
Measured & 0.27 & 0.03 & -0.54 & 0.17 & -0.19 & 0.04\\
\hline\\
& $\mu^{142}\rm Nd$ & $\pm$ & $\varepsilon^{64}\rm Ni$ & $\pm$ & $\varepsilon^{92}\rm Mo$ & $\pm$\\
\hline\\  
Calculated & -12.5 & 2.6 & -0.07 & 0.11 & 0.63 & 0.21\\
Measured & -16 & 8 & 0.10 & 0.28 & 0.20 & 0.53 \\
\hline
\end{tabular}
%\tablecomments{}
\end{table*} 

We adopt a different method in calculating the $\varepsilon_{\rm Mars}$ for $^{64} \rm Ni$ and $^{92} \rm Mo$. Both Ni and Mo are moderately siderophile elements and behave more like siderophiles than Cr. It is probable that a significant portion of accreted Ni and Mo sank into the core throughout the accretion of Mars. \cite{dauphas2017isotopic} showed that nearly all the Ni and Mo accreted in the first stage of Earth’s accretion should have sunk into the core. The first stage of Earth’s accretion is defined as the first 60$\%$ of Earth’s accretion in terms of mass because previous models suggested that material accreted by Earth changed from more reduced to more oxidised at about 60$\%$ accretion \citep{rubie2015accretion}. It is expected that Ni and Mo should behave similarly in Mars. For Ni and Mo, we ignore all accretion during the first $\sim$60\% of Mars’ accretion. This requires a higher resolution of the late-stage accretion data for each Mars analogue, which our simulations lack. Therefore, we can only pick the data from simulations with smaller initial embryo mass (0.025 $M_{\Earth}$). The average percentage of enstatite, ordinary chondrite, and carbonaceous chondrite is calculated for 20 Mars analogues from their last $\sim$40\% of accretion. The $\varepsilon_{\rm Mars}$ value of $^{64} \rm Ni$ and $^{92} \rm Mo$ are calculated with the same Monte Carlo method described above, but with the $f_i$ obtained from the last $\sim$40\% of accretion. The mean bulk composition ($\pm$1$\sigma$) from $N$-body simulations is $41\%^{+23}_{-23}$ enstatite chondrite, $58\%^{+24}_{-24}$ ordinary chondrite, and $0.8\%^{+3.6}_{-0.8}$ carbonaceous
chondrite for the last $\sim$40\% of Mars' accretion when the break location of the disk is at 1.3 AU. Table \ref{tb:isotope} shows that the calculated values match with the measured values for both Ni and Mo. 

%Results for $^{92} \rm Mo$ agree within the huge error bars but our calculated nominal value is about 3 times larger than the measured nominal value. One of the possible reasons is that we only include chondrite when calculating $\varepsilon_{\rm Mars}$. \textbf{The nominal $\varepsilon^{92} \rm Mo$ of enstatite, ordinary and carbonaceous chondrite are all significantly higher than the measured $\varepsilon_{\rm Mars}$ \citep{burkhardt2011molybdenum,burkhardt2014evidence,dauphas2002inference,dauphas2002molybdenum,dauphas2017isotopic}.  Including some other type of meteorite is one of possible ways to lower the calculated $\varepsilon_{\rm Mars}$ of $^{92} \rm Mo$. Perhaps achondrite such as the angrites, aubrites or ureilites, which experienced melting and recrystallisation within their parent bodies, is part or all of the missing extra component. The angrites have a negative value for $\varepsilon^{92} \rm Mo$, but with large uncertainty \citep{burkhardt2011molybdenum}. This may further indicate that Mars accreted more achondrite in its later stages of accretion. Better measurement for the $\varepsilon_{\rm Mars}$ of $^{92} \rm Mo$ from martian meteorites and the bulk Mo concentration of the angrites will definely help making more explicit conclusion}.

To summarise, it is possible to reproduce the isotopic composition of the martian mantle for six isotopes from our simulation with the Grand Tack model. The uncertainties of the martian isotopic composition calculated from the $N$-body simulations for 4 out of 6 elements are of the same order of magnitude as the measured data. This is caused mainly by the great variety of Mars' dynamical evolution in our $N$-body simulations. Therefore, to better constrain the outcome, more high-resolution simulations are needed. We encourage improved measurements of the isotopic composition of martian meteorites to lower the measured uncertainties.

\section{Conclusions}
\label{sec:conclusions}
The isotopic anomaly between Earth and Mars in several nuclides suggests that the bulk compositions of Earth and Mars are different. Assuming they both accreted mainly chondritic material, Earth should have inherited mainly enstatite chondrite ($\sim$70\%) and Mars is likely to have accreted more ordinary chondrite (32\% to 67\%) than Earth. This provides further support to the conclusion that the region where Mars accreted most of its mass was different from that of Earth, and was likely to have been more distant. We first examine the possibility of forming Earth and Mars with different compositions in two different terrestrial planet formation models: the Grand Tack model and the Classical model. By tracing the initial positions of the planetesimals and embryos assembled into each planet, an accretion zone comparison can be made between the Earth and Mars analogues after performing a large number of $N$-body simulations. Our initial conditions are either embryos with equal masses or with masses that were computed from the traditional oligarchic growth of the planetesimals. We find that there are seven Mars analogues with $a_{\rm mwmi}$ $>$ 2 AU in the Grand Tack simulations. These are Mars analogues that have a very different accretion zone from all the Earth analogues. They only contribute $<$ 5$\%$ of the total Mars analogues. We discovered no Mars analogues with $a_{\rm mwmi}$ $>$ 2 AU in the Classical model when we considered the massive Mars problem generally observed in the Classical model's simulations. Both the Classical and Grand Tack have difficulty in explaining the compositional difference between Earth and Mars. We also find that the final number of Mars analogues we obtain from the simulations is independent of the initial number of embryos, which violates our original expectation that more initial embryos yield more Mars analogues.

We then estimated the average percentage of enstatite chondrite and ordinary chondrite incorporated into Earth and Mars based on the $N$-body simulation results. At first we assumed that the solid disk is made entirely of enstatite and ordinary chondrite, with enstatite chondrite residing in the inner region and ordinary chondrite in the outer region. The break location of the disk, at which the disk changes from enstatite chondrite to ordinary chondrite, is set as a free parameter from 1 to 2 AU with 0.1 AU intervals. Our results show that the break location must be close to 1.3 AU for both models to possess mean compositions of Earth and Mars close to the documented ones. At this break location, the Classical model yields more Mars analogues close to Mars' documented composition than the Grand Tack model if we neglect the problem of excessive mass of Mars in the Classical model. This composition requires an early collision between two embryos of different composition that merge together to form Mars, which agrees with the distant Mars formation scenario \citep{brasser2017cool}.

In the previous section, we calculated the isotopic composition from our simulation results for four elements that trace the full accretion history of Mars ($^{17} \rm O$, $^{50} \rm Ti$, $^{54} \rm Cr$ and $^{142} \rm Nd$) and two elements that trace only the late accretion of Mars ($^{64} \rm Ni$ and $^{92} \rm Mo$) in the case of the break location of the disk at 1.3 AU, in which Mars is $\sim$50\% enstatite chondrite and $\sim$50\% ordinary chondrite. The Grand Tack model matches the measured values within their uncertainties for all elements. Nevertheless, we cannot at the present time confine the dynamical pathway of Mars by matching the calculated isotopic composition to the measured values since their uncertainties are comparable to each other. More dynamical simulations on terrestrial planet formation with higher resolution are needed. We further advocate improved measurements of the isotopic composition of martian meteorites in order to solve the mystery of the formation of Mars and the other terrestrial planets.

\begin{acknowledgements}
RB is grateful for financial support from JSPS KAKENHI (JP16K17662). RB and SJM acknowledge the Collaborative for Research in Origins (CRiO), which is supported by The John Templeton Foundation – FfAME Origins program: the opinions expressed in this publication are those of the authors, and do not necessarily reflect the views of the John Templeton Foundation. The source codes for the model used in this study are archived at the Earth Life Science Institute of the Tokyo Institute of Technology. The data, input, and output files necessary to reproduce the figures are available from the authors upon request.
\end{acknowledgements}

\bibliographystyle{aa} 
\bibliography{woo_etal_2018_final}

\newcommand{\noop}[1]{}
\begin{thebibliography}{52}
\expandafter\ifx\csname natexlab\endcsname\relax\def\natexlab#1{#1}\fi

\bibitem[{Agee {et~al.}(2013)Agee, Wilson, McCubbin, Ziegler, Polyak, Sharp,
  Asmerom, Nunn, Shaheen, Thiemens, {et~al.}}]{agee2013unique}
Agee, C.~B., Wilson, N.~V., McCubbin, F.~M., {et~al.} 2013, Science, 339, 780

\bibitem[{Agnor {et~al.}(1999)Agnor, Canup, \& Levison}]{agnor1999character}
Agnor, C.~B., Canup, R.~M., \& Levison, H.~F. 1999, Icarus, 142, 219

\bibitem[{Bitsch {et~al.}(2015)Bitsch, Johansen, Lambrechts, \&
  Morbidelli}]{bitsch2015structure}
Bitsch, B., Johansen, A., Lambrechts, M., \& Morbidelli, A. 2015, Astronomy \&
  Astrophysics, 575, A28

\bibitem[{Bitsch {et~al.}(2014)Bitsch, Morbidelli, Lega, Kretke, \&
  Crida}]{bitsch2014stellar}
Bitsch, B., Morbidelli, A., Lega, E., Kretke, K., \& Crida, A. 2014, Astronomy
  \& Astrophysics, 570, A75

\bibitem[{Brasser {et~al.}(2018)Brasser, Dauphas, \& Mojzsis}]{brasser2018GRL}
Brasser, R., Dauphas, N., \& Mojzsis, S.~J. 2018, Geophysical Research Letter,
  in press

\bibitem[{Brasser {et~al.}(2016)Brasser, Matsumura, Ida, Mojzsis, \&
  Werner}]{brasser2016analysis}
Brasser, R., Matsumura, S., Ida, S., Mojzsis, S., \& Werner, S. 2016, The
  Astrophysical Journal, 821, 75

\bibitem[{Brasser {et~al.}(2017)Brasser, Mojzsis, Matsumura, \&
  Ida}]{brasser2017cool}
Brasser, R., Mojzsis, S., Matsumura, S., \& Ida, S. 2017, Earth and Planetary
  Science Letters, 468, 85

\bibitem[{Chambers(2001)}]{chambers2001making}
Chambers, J. 2001, Icarus, 152, 205

\bibitem[{Chambers(2006)}]{chambers2006semi}
Chambers, J. 2006, Icarus, 180, 496

\bibitem[{Coradini {et~al.}(2011)Coradini, Capaccioni, Erard, Arnold,
  De~Sanctis, Filacchione, Tosi, Barucci, Capria, Ammannito,
  {et~al.}}]{coradini2011surface}
Coradini, A., Capaccioni, F., Erard, S., {et~al.} 2011, Science, 334, 492

\bibitem[{Dauphas(2017)}]{dauphas2017isotopic}
Dauphas, N. 2017, Nature, 541, 521

\bibitem[{Dauphas {et~al.}(2014)Dauphas, Chen, Zhang, Papanastassiou, Davis, \&
  Travaglio}]{dauphas2014calcium}
Dauphas, N., Chen, J.~H., Zhang, J., {et~al.} 2014, Earth and Planetary Science
  Letters, 407, 96

\bibitem[{Dauphas \& Pourmand(2011)}]{dauphas2011hf}
Dauphas, N. \& Pourmand, A. 2011, Nature, 473, 489

\bibitem[{DeMeo \& Carry(2014)}]{demeo2014solar}
DeMeo, F. \& Carry, B. 2014, Nature, 505, 629

\bibitem[{Dr{\k{a}}{\.z}kowska {et~al.}(2016)Dr{\k{a}}{\.z}kowska, Alibert, \&
  Moore}]{drkazkowska2016close}
Dr{\k{a}}{\.z}kowska, J., Alibert, Y., \& Moore, B. 2016, Astronomy \&
  Astrophysics, 594, A105

\bibitem[{Duncan {et~al.}(1998)Duncan, Levison, \& Lee}]{duncan1998multiple}
Duncan, M.~J., Levison, H.~F., \& Lee, M.~H. 1998, The Astronomical Journal,
  116, 2067

\bibitem[{Fischer {et~al.}(2018)Fischer, Nimmo, \& O'Brien}]{fischer2018radial}
Fischer, R.~A., Nimmo, F., \& O'Brien, D.~P. 2018, Earth and Planetary Science
  Letters, 482, 105

\bibitem[{Fischer-G{\"o}dde \& Kleine(2017)}]{fischer2017ruthenium}
Fischer-G{\"o}dde, M. \& Kleine, T. 2017, Nature, 541, 525

\bibitem[{Fitoussi {et~al.}(2016)Fitoussi, Bourdon, \&
  Wang}]{fitoussi2016building}
Fitoussi, C., Bourdon, B., \& Wang, X. 2016, Earth and Planetary Science
  Letters, 434, 151

\bibitem[{Franchi {et~al.}(1999)Franchi, Wright, Sexton, \&
  Pillinger}]{franchi1999oxygen}
Franchi, I., Wright, I., Sexton, A., \& Pillinger, C. 1999, Meteoritics \&
  Planetary Science, 34, 657

\bibitem[{Gradie \& Tedesco(1982)}]{gradie1982compositional}
Gradie, J. \& Tedesco, E. 1982, Science, 216, 1405

\bibitem[{Hansen(2009)}]{hansen2009formation}
Hansen, B.~M. 2009, The Astrophysical Journal, 703, 1131

\bibitem[{Hayashi(1981)}]{hayashi1981structure}
Hayashi, C. 1981, Progress of Theoretical Physics Supplement, 70, 35

\bibitem[{Ida \& Guillot(2016)}]{ida2016formation}
Ida, S. \& Guillot, T. 2016, Astronomy \& Astrophysics, 596, L3

\bibitem[{Jacobson \& Morbidelli(2014)}]{jacobson2014lunar}
Jacobson, S.~A. \& Morbidelli, A. 2014, Phil. Trans. R. Soc. A, 372, 20130174

\bibitem[{Johansen {et~al.}(2007)Johansen, Oishi, Mac~Low, Klahr, Henning, \&
  Youdin}]{johansen2007rapid}
Johansen, A., Oishi, J.~S., Mac~Low, M.-M., {et~al.} 2007, Nature, 448, 1022

\bibitem[{Kaib \& Cowan(2015)}]{kaib2015feeding}
Kaib, N.~A. \& Cowan, N.~B. 2015, Icarus, 252, 161

\bibitem[{Kleine {et~al.}(2009)Kleine, Touboul, Bourdon, Nimmo, Mezger, Palme,
  Jacobsen, Yin, \& Halliday}]{kleine2009hf}
Kleine, T., Touboul, M., Bourdon, B., {et~al.} 2009, Geochimica et Cosmochimica
  Acta, 73, 5150

\bibitem[{Kokubo \& Ida(1998)}]{kokubo1998oligarchic}
Kokubo, E. \& Ida, S. 1998, Icarus, 131, 171

\bibitem[{Kruijer {et~al.}(2017)Kruijer, Kleine, Borg, Brennecka, Irving,
  Bischoff, \& Agee}]{kruijer2017early}
Kruijer, T.~S., Kleine, T., Borg, L.~E., {et~al.} 2017, Earth and Planetary
  Science Letters, 474, 345

\bibitem[{Lin \& Papaloizou(1986)}]{lin1986tidal}
Lin, D. \& Papaloizou, J. 1986, The Astrophysical Journal, 309, 846

\bibitem[{Lodders(2000)}]{lodders2000oxygen}
Lodders, K. 2000, in From Dust to Terrestrial Planets (Springer), 341--354

\bibitem[{Masset \& Snellgrove(2001)}]{masset2001reversing}
Masset, F. \& Snellgrove, M. 2001, Monthly Notices of the Royal Astronomical
  Society, 320, L55

\bibitem[{Mittlefehldt {et~al.}(2008)Mittlefehldt, Clayton, Drake, \&
  Righter}]{mittlefehldt2008oxygen}
Mittlefehldt, D.~W., Clayton, R.~N., Drake, M.~J., \& Righter, K. 2008, Reviews
  in Mineralogy and Geochemistry, 68, 399

\bibitem[{Morbidelli {et~al.}(2012)Morbidelli, Lunine, O'Brien, Raymond, \&
  Walsh}]{morbidelli2012building}
Morbidelli, A., Lunine, J.~I., O'Brien, D.~P., Raymond, S.~N., \& Walsh, K.~J.
  2012, Annual Review of Earth and Planetary Sciences, 40, 251

\bibitem[{O'Brien {et~al.}(2018)O'Brien, Izidoro, Jacobson, Raymond, \&
  Rubie}]{o2018delivery}
O'Brien, D.~P., Izidoro, A., Jacobson, S.~A., Raymond, S.~N., \& Rubie, D.~C.
  2018, arXiv preprint arXiv:1801.05456

\bibitem[{Ogihara \& Ida(2009)}]{ogihara2009n}
Ogihara, M. \& Ida, S. 2009, The Astrophysical Journal, 699, 824

\bibitem[{Qin {et~al.}(2010)Qin, Alexander, Carlson, Horan, \&
  Yokoyama}]{qin2010contributors}
Qin, L., Alexander, C.~M., Carlson, R.~W., Horan, M.~F., \& Yokoyama, T. 2010,
  Geochimica et Cosmochimica Acta, 74, 1122

\bibitem[{Qin \& Carlson(2016)}]{qin2016nucleosynthetic}
Qin, L. \& Carlson, R.~W. 2016, Geochemical Journal, 50, 43

\bibitem[{Raymond {et~al.}(2009)Raymond, O’Brien, Morbidelli, \&
  Kaib}]{raymond2009building}
Raymond, S.~N., O’Brien, D.~P., Morbidelli, A., \& Kaib, N.~A. 2009, Icarus,
  203, 644

\bibitem[{Righter \& Chabot(2011)}]{righter2011moderately}
Righter, K. \& Chabot, N.~L. 2011, Meteoritics \& Planetary Science, 46, 157

\bibitem[{Rubie {et~al.}(2015)Rubie, Jacobson, Morbidelli, O’Brien, Young,
  de~Vries, Nimmo, Palme, \& Frost}]{rubie2015accretion}
Rubie, D.~C., Jacobson, S.~A., Morbidelli, A., {et~al.} 2015, Icarus, 248, 89

\bibitem[{Rubin {et~al.}(2000)Rubin, Warren, Greenwood, Verish, Leshin, Hervig,
  Clayton, \& Mayeda}]{rubin2000angeles}
Rubin, A.~E., Warren, P.~H., Greenwood, J.~P., {et~al.} 2000, Geology, 28, 1011

\bibitem[{Sanloup {et~al.}(1999)Sanloup, Jambon, \& Gillet}]{sanloup1999simple}
Sanloup, C., Jambon, A., \& Gillet, P. 1999, Physics of the Earth and Planetary
  Interiors, 112, 43

\bibitem[{Tang \& Dauphas(2014)}]{tang201460fe}
Tang, H. \& Dauphas, N. 2014, Earth and Planetary Science Letters, 390, 264

\bibitem[{Vernazza {et~al.}(2009)Vernazza, Brunetto, Binzel, Perron, Fulvio,
  Strazzulla, \& Fulchignoni}]{vernazza2009plausible}
Vernazza, P., Brunetto, R., Binzel, R., {et~al.} 2009, Icarus, 202, 477

\bibitem[{Vernazza {et~al.}(2011)Vernazza, Lamy, Groussin, Hiroi, Jorda, King,
  Izawa, Marchis, Birlan, \& Brunetto}]{vernazza2011asteroid}
Vernazza, P., Lamy, P., Groussin, O., {et~al.} 2011, Icarus, 216, 650

\bibitem[{Walsh {et~al.}(2011)Walsh, Morbidelli, Raymond, O'brien, \&
  Mandell}]{walsh2011low}
Walsh, K.~J., Morbidelli, A., Raymond, S.~N., O'brien, D.~P., \& Mandell, A.~M.
  2011, Nature, 475, 206

\bibitem[{W{\"a}nke \& Dreibus(1988)}]{wanke1988chemical}
W{\"a}nke, H. \& Dreibus, G. 1988, Phil. Trans. R. Soc. Lond. A, 325, 545

\bibitem[{W{\"a}nke \& Dreibus(1994)}]{wanke1994chemistry}
W{\"a}nke, H. \& Dreibus, G. 1994, Phil. Trans. R. Soc. Lond. A, 349, 285

\bibitem[{Warren(2011)}]{warren2011stable}
Warren, P.~H. 2011, Earth and Planetary Science Letters, 311, 93

\bibitem[{Wittmann {et~al.}(2015)Wittmann, Korotev, Jolliff, Irving, Moser,
  Barker, \& Rumble}]{wittmann2015petrography}
Wittmann, A., Korotev, R.~L., Jolliff, B.~L., {et~al.} 2015, Meteoritics \&
  Planetary Science, 50, 326

\end{thebibliography}

\end{document}